\documentclass[twocolumn,showpacs,preprintnumbers,amsmath,amssymb]{revtex4}
\usepackage{epsfig}

\usepackage{mathptmx}				
\renewcommand*{\vec}{\mathbf}			

\begin{document}

\title{Neutrino orbital angular momentum in a plasma vortex}

\author{J.\,T. Mendon\c{c}a}

\email{titomend@ist.utl.pt}

\affiliation{IPFN and CFIF, Instituto Superior T\'{e}cnico, 1049-001 Lisboa,
Portugal}

\author{B. Thid\'e}

\altaffiliation[Also at ]{LOIS Space Centre, V\"axj\"o University,
SE-351\,95 V\"axj\"o, Sweden}
\affiliation{Swedish Institute of Space Physics, P.\,O. Box 537,
SE-751\,21 Uppsala, Sweden}

\begin{abstract}

It is shown that an electron-neutrino beam, propagating in a background plasma, can be decomposed into orbital momentum (OAM) states, similar to the OAM photon states. Coupling between different OAM neutrino states, in the presence of a plasma vortex, is considered. We show that plasma vorticity can be transfered to the neutrino beam, which is relevant to the understanding of the neutrino sources in astrophysics. Observation of neutrino OAM states could eventually become possible.

\end{abstract}

\maketitle


\section{Introduction}

Neutrino beam interaction with a dense plasma has received considerable
attention in recent years.  In particular, it was recognized that
collective neutrino-plasma displays strong similarities with
laser-plasma interaction phenomena.  This is due to the fact that
neutrino dispersion relation in a plasma is formally analogous to the
photon dispersion relation \cite{bethe}.  Coupling between neutrino
beams and dense plasmas can then be conceived \cite{bingham94,mend95},
which is particularly relevant for the understanding of supernova type
II explosions, where an intense and short neutrino burst is emitted
\cite{bethewilson}.  It was also recognized that neutrino ponderomotive
force can excite kinetic plasma instabilities \cite{silva,oraevsky},
resulting from negative neutrino Landau damping of electron plasma
waves, similar to that observed for photons \cite{bobmend97}.
Furthermore, it is also possible to show that neutrinos acquire an
effective electric charge in a plasma, in a way very similar to photons
\cite{serbeto,mend02}.  Another important aspect is related to the
possible excitation of relativistic plasma wakefields by a neutrino
burst \cite{padma,rios}, and the generation of magnetic fields
\cite{padmabis}, through processes that mimic those of photon beams in a
plasma.

In this work, the similarities between neutrino and photon dispersion
are explored even further, by introducing the concept of neutrino
orbital angular momentum (OAM).  This is a new aspect of neutrino plasma
physics, which is related with the possible exchange of orbital angular
momentum between a neutrino beam and a rotating plasma.  It is known
that photons can carry, not only intrinsic angular momentum or spin,
which is associated with their polarization state, but also orbital
angular momentum (OAM) \cite{allen}.  The existence of OAM photon states
has been experimentally demonstrated early in 1936 \cite{beth36}, but
only recently photon OAM momentum received attention, after the
demonstration that Laguerre-Guassian laser modes correspond to well
defined OAM modes, and that these photon modes not only can be measured
as a photon beam property \cite{harris,padgett}, but can also be
detected at the single photon level \cite{leach}.

Utilization of photon OAM in the low frequency radio wave domain was
recently proposed by one of us \cite{thide}, as an additional method for
studying the properties of radio sources.  Here we show that a neutrino
beam can be decomposed into OAM momentum states, similar to the OAM
photon states, and we study the coupling between these modes, when the
neutrino beam propagates across a plasma vortex region.  We will also
show that plasma vorticity can be transferred to the neutrino beam,
which can likewise be of great relevance to the understanding of the
astrophysical sources of neutrino emission.  Detection of finite
neutrino OAM states would give us additional information on the
collapsing star, in the case of a supernova explosion, and would allow
us to measure the vortex properties of the surrounding dense plasma.
Observation of neutrino OAM states could eventually be possible, by
studying the conservation of total angular momentum at detection, and by
adapting the existing detection schemes to that purpose.  Detection of
radio signals from neutrinos interacting with the Moon, as recently
proposed \cite{stal07}, could provide one possible method for measur ing
such neutrino OAM states.

\section{Basic equations}

We know that the spinor field $ \psi $ describing electron-neutrinos moving in a dense plasma can be described by the Dirac equation
\begin{equation}
i \hbar \frac{\partial}{\partial t}  \psi  = H  \psi   \quad , \quad H = \beta m_\nu c^2 + \vec{\alpha} \cdot \vec{p} c + V
\label{eq:2.1} \end{equation}
where $m_\nu$ is a non-zero rest mass, $\beta$ and $\vec{\alpha}$ are the well known operators.  For simplicity, the plasma medium is assumed isotropic, and magnetic field effects are ignored. 
We retain explicit physical units, with $c \neq 1$ and $\hbar \neq 1$, for future comparison between neutrino and photon states in a plasma.
The effective plasma potential $V (\vec{r}, t)$, results from the weak interactions between the neutrinos and the background electrons and ions (or protons) of the background plasma. We know that, for an electrically neutral medium, the proton contributions are exactly canceled by the neutral part of the electron contribution, and the resulting plasma potential is simply given by 
\begin{equation}
V  = \sqrt{2} G_F n_e (\vec{r}, t) = \hbar g \omega_p^2  \quad , \quad g = \sqrt{2} \frac{\epsilon_0 m_e}{e^2 \hbar} G_F
\label{eq:2.2} \end{equation}
where $n_e$ is the electron mean plasma density, $\omega_p$ is the electron plasma frequency, $e$ and $m_e$ are the electron charge and mass, $\epsilon_0$ is the permittivity of vacuum, and $G_F$ is the Fermi constant. Equations (\ref{eq:2.1})-(\ref{eq:2.2}) have been used to study the collective decay of plasmons into neutrino-antineutrino pairs \cite{serb,urca}. Here, we neglect such neutrino-antineutrino coupling processes. We also neglect mass oscillations,  but their possible contribution will be discussed later. For a constant plasma density, we can use the Foldy-Wouthuysen procedure, leading to a diagonalized Hamiltonian
\begin{equation}
H = \beta \sqrt{m_\nu^2 c^4 + p^2 c^2} + V
\label{eq:2.3} \end{equation}
We will focus on a single spin (or helicity, in the limit of a negligible mass) electron-neutrino state. The corresponding solution can then be written as
$\psi (\vec{r}, t) = \Psi_0 \exp( i \vec{k} \cdot \vec{r} - i \omega t)$,
where the neutrino frequency $\omega = < H > / \hbar$, and the wavevector $\vec{k} = < \vec{p} > / \hbar$ satisfy the neutrino dispersion relation
\begin{equation}
(\omega - g \omega_p^2)^2 = k^2 c^2 + \omega_0^2
\label{eq:2.4} \end{equation}
with $\omega_0 = m_\nu c^2 / \hbar$. This shows a strong similarity with the photon dispersion relation in a plasma.
Let us now consider a non-homogeneous  and non-stationary medium, where the plasma density and the plasma potential are no longer constant, but vary on time a length scales much longer than the neutrino period and wavelength. 
\begin{equation}
V (\vec{r}, t) = V_0 + \tilde{V} (\vec{r}, t) 
\label{eq:3.1} \end{equation}
which is the consequence of a space-time variation of the background plasma density and of the corresponding plasma frequency, as described by
\begin{equation}
\omega_p^2 = \omega_{p0}^2 [ 1 + \epsilon (\vec{r}, t) ] = \omega_{p0}^2 \left[ 1 + \frac{1}{n_0} \tilde{n} (\vec{r}, t) \right]
\label{eq:3.2} \end{equation}
where $\tilde{n}$ is the electron density perturbation. By comparing the two expressions, we conclude that
\begin{equation}
\tilde{V} (\vec{r}, t) = \hbar g \omega_{p0}^2 \epsilon (\vec{r}, t) = \hbar g \frac{\omega_{p0}^2}{n_0} \tilde{n} (\vec{r}, t) 
\label{eq:3.2b} \end{equation}
In order to solve the neutrino field equation, we can use the quasi-classical or WKB approximation and make the replacements
\begin{equation}
H \; \rightarrow \; \hbar \left( \omega - i \frac{\partial}{\partial t} \right) \quad , \quad
\vec{p} \; \rightarrow \; \hbar (\vec{k} + i \nabla). 
\label{eq:2.5} \end{equation} 
The Hamiltonian operator (\ref{eq:2.3}) then becomes
\begin{equation}
\left( \omega - i \frac{\partial}{\partial t} \right) = \beta \sqrt{\omega_0^2 + (\vec{k} + i \nabla)^2 c^2} + \frac{1}{\hbar} (V_0 + \tilde{V})
\label{eq:2.5b} \end{equation}
Assuming that the dispersion relation (\ref{eq:2.4}) is always locally satisfied for the non-perturbed plasma potential $V_0$, we can use a WKB solution of the form
\begin{equation}
\psi (\vec{r}, t) = \Psi (\vec{r}, t) \exp \left( i \vec{k} \cdot \vec{r} - i \omega  t \right)
\label{eq:2.6} \end{equation}
where $\Psi (\vec{r}, t)$ is now a slowly varying amplitude. Expanding the operator (\ref{eq:2.5}) we can then write the evolution equation for this amplitude as
\begin{equation}
- i \hbar \frac{\partial}{\partial t} \Psi = \frac{c^2}{(\omega - g \omega_p^2)} \left[ i \vec{k} \cdot \nabla - \frac{1}{2} \nabla^2 - \frac{c^2(\vec{k} \cdot \nabla)^2}{2 (\omega - g \omega_p^2)} + \tilde{V} \right] \Psi 
\label{eq:2.7} \end{equation}
Noting that the neutrino mean velocity can be identified with the group velocity associated with the dispersion relation (\ref{eq:2.4}), we get
\begin{equation}
\vec{v}_\nu = \frac{\partial \omega}{\partial \vec{k}} = \frac{c^2 \vec{k}}{\omega - g \omega_p^2}
\label{eq:2.7b} \end{equation}
We can also introduce the neutrino relativistic factor $\gamma_\nu$, as defined by
\begin{equation}
m_\nu \gamma_\nu c^2 = \hbar (\omega - g \omega_p^2)
\label{eq:2.7c} \end{equation}
Using these quantities, we can rewrite equation (\ref{eq:2.7}) in a more appropriate form
\begin{equation}
i \left( \frac{\partial}{\partial t} + \vec{v}_\nu \cdot \nabla \right) \Psi = \frac{\hbar}{2 m_\nu \gamma_\nu} \left[ \nabla^2 + \left( \frac{\vec{v}_\nu}{c} \cdot \nabla \right)^2  + g \frac{\omega_{p0}^2}{n_0} \tilde{n}  \right] \Psi
\label{eq:2.7d} \end{equation}
In the limit of a negligible neutrino velocity, this would take the form of a Schroedinger equation. A similar equation was also derived for the case of a quantum free electron laser \cite{fel,prep}. 
Let us first assume that a neutrino beam propagates along a given axis $Oz$, in the absence of any plasma perturbation, $\tilde{n} = 0$. Retaining only the first order derivatives in the z- direction, we can then reduce the wave equation to a paraxial equation, similar to that describing a laser beam in the focal region
\begin{equation}
\left( \nabla_\perp^2 + 2 i k  \frac{\partial \psi}{\partial z}  \right) \psi_0 = 0
\label{eq:2.8} \end{equation}
where the wavenumber $k$ obeys the dispersion relation (\ref{eq:2.4}). We can try a Gaussian solution of the form
\begin{equation}
\psi_0 (\vec{r}) \equiv \psi_0 (r, z) = \Psi_0 (z) \exp \left[ i k \frac{r^2}{2 R (z)} \right]
\label{eq:2.9} \end{equation}
Using the transverse Laplacian in cylindrical coordinates, we realize that this solution satisfies the paraxial equation, if we assume that
\begin{equation}
R (z) = R_0 + (z - z_0) \quad , \quad
\Psi_0 (z) = u_0 \frac{R_0}{R (z)}
\label{eq:2.10} \end{equation}
where $R_0$ and $u_0$ are constants. Very simple arguments can be used to justify the relevance of such Gaussian beam solutions in neutrino physics. For instance, the intense neutrino burst emitted by a collapsing type II supernova, is naturally focused by the plasma inhomogeneities of the star material. The neutrino dispersion relation (\ref{eq:2.4}) shows that neutrinos will be focused into the lower density plasma regions, in the same way as high energy photons would do. Another reason for neutrino focusing into a nearly Gaussian form, would be the occurrence of filamentational instability of the ultra intense neutrino burst coming out of the collapsing star.
A more general cylindrical type of solution of the paraxial equation (\ref{eq:2.8}) can be represented in the basis of orthogonal Laguerre-Gaussian modes, as defined by
\begin{equation}
\psi (\vec{r}, t) =  \sum_{pl} \;  \Phi_{pl}F_{pl} (r, \phi) \exp (i k z - i \omega t)
\label{eq:2.11} \end{equation}
with
\begin{equation}
F_{pl} (r, \phi) \; \propto \;  \left( \frac{r^2}{w^2} \right)^{|l|} L_p^{|l|} \left( \frac{r^2}{w^2} \right) \exp( i l \phi - \frac{r^2}{2 w^2} )
\label{eq:2.12} \end{equation}
The relevance of this type of representation will become apparent below. We recover the particular Gaussian solution by taking $l = 0$, and $R = - i w^2$. The new quantity $w$ defines the neutrino beam waist.
By making an appropriate choice of the normalization factors multiplying equation (\ref{eq:2.12}), we can verify the orthogonality conditions
\begin{equation}
\int_0^\infty r dr \int_0^{2 \pi} d \phi F_{pl}^* (r, \phi) F_{p'l'} (r, \phi) = \delta_{pp'} \delta_{ll'}
\label{eq:2.13} \end{equation}

\section{Neutrino mode coupling}

Let us now consider the case where the electron density and the plasma potential are not constant, $\tilde{n} \neq 0$. In cylindrical geometry, we can use a general representation in terms of the Laguerre-Gaussian modes, as 
\begin{equation}
\tilde{n} (\vec{r}, t)  = \sum_l \int \frac{d q}{2 \pi}  \int \frac{d \Omega}{2 \pi} \tilde{n}_l (r, q, \Omega) \exp (i l \phi + i q z - i \Omega t) 
\label{eq:3.3} \end{equation}
To be more specific, we can concentrate on helical type of perturbations described by
\begin{equation} 
\tilde{n} (\vec{r}, t)  = \tilde{n} (r) \exp \left( - \frac{r^2}{2 a^2} \right)  \cos (l_0 \phi + i q z - i \Omega t) 
\label{eq:3.4} \end{equation}
In the presence of such density perturbations, the different Laguerre-Gaussian modes of the neutrino field will be coupled, and the generic wavefunction solution will take the form
\begin{equation}
\psi (\vec{r}, t) = \sum_{n,p,l} \psi_{npl} (\vec{r}) \exp [i S (\vec{r}, t)]
\label{eq:3.5} \end{equation}
where $\psi_{npl}$ are slowly varying ampitudes, and $S (\vec{r}, t)$ is the phase function as defined by
$S (\vec{r}, t) = k_n z - \omega_n t$. Here we have used $\omega_n = \omega + n \Omega$, and the corresponding wavenumbers $k_n$ are determined by the neutrino dispersion, with $\omega$ replaced by $\omega_n$, valid for the unperturbed plasma density $n_0$. We can use, for the above amplitudes 
\begin{equation}
\psi_{npl} (\vec{r}) = \Psi_{npl} (z) \; F_{pl} (r, \phi)
\label{eq:3.5b} \end{equation}
Replacing this in the paraxial equation, where the term corresponding to the perturbed potential is retained, we obtain the following mode coupling equations
\begin{equation}
\frac{d}{d z} \Psi_{npl} (z) = - i \sum_{n'p'l'} K (npl,n'p'l') \Psi_{n'p'l'} (z)
\exp ( i \Delta_{nn'} z)
\label{eq:3.6} \end{equation}
where we have defined the wavenumber mismatch $\Delta_{nn'} = (k_{n'} - k_n \pm q)$, and  used the following coupling coefficients
\begin{multline}\label{eq:3.7}
K (npl,n'p'l') = \frac{g \omega_n}{k_n c^2}
 \frac{\omega_{p0}^2}{n_0} \delta (n - n' \pm 1) \\
 \times\int_0^{2 \pi} d \phi \int_0^\infty r dr \; \tilde{n} (r) F_{p'l'} (r) F_{pl}^* (r) \exp[ i ( l' \pm l_0 - l) \phi]
\end{multline}
Here, in the expressions of $F_{pl}^* (r)$ and $F_{p'l'} (r)$ we have taken out the factors $\exp (- i l \phi)$ and $\exp (i l' \phi)$. The coupled mode equation (\ref{eq:3.6}) shows that, if we start from an initial neutrino state with no orbital angular momentum, $l' = 0$, we will excite states of finite angular momentum $l \neq 0$, due to the existence of a perturbed plasma helical structure. But such coupling has to verify certain resonant conditions. Depending on the radial profile of the plasma vortex $\tilde{n} (r)$, various modes $p \neq p'$ can be excited. But we take here the simplest case of $p = p'$, which allows us to drop this index from the coupled mode equations,  and write them in a much simpler form
\begin{equation}
\frac{d}{d z} \Psi_\nu (z) = - i  K \left[ \Psi_{\nu+1} (z) e^{ i \Delta z} + \Psi_{\nu-1} (z) e^{- i \Delta z} \right]
\label{eq:3.8} \end{equation}
Here $\nu$ is an integer, and we have used a simplified notation for the amplitudes, $\Psi_\nu \equiv \Psi_{n+\nu,p,l+\nu l_0}$. Approximate expressions for the coupling coefficients and wavenumber mismatch can be written as
\begin{equation}
K \simeq \frac{g \omega}{k c^2} \frac{\omega_{p0}^2}{n_0} \tilde{n} (0) \quad , \quad \Delta \simeq \left( \frac{\Omega}{c} + q \right)
\label{eq:3.8b} \end{equation}
For an exact phase matching, such that $q = - \Omega / c$, the coupled mode equations (\ref{eq:3.8}) can easily be integrated in terms of Bessel functions. For initial conditions such that $\Psi_l (0) = \Psi_0 \delta_{l0}$, corresponding to an initial Gaussian neutrino beam, we get the following amplitude for the various interacting modes
\begin{equation}
\Psi_\nu (z) = i^\nu \Psi_0 J_\nu ( 2 K z)
\label{eq:3.9} \end{equation}
This shows that the states of higher neutrino OAM can be populated by a linear cascading process, approximately described by this simple law. Such a simple but physically relevant solution is valid for arbitrary interacting distances in the case of a perfect phase matching $\Delta = 0$. And it remains valid for a finite phase mismatch, for interaction distances much shorter that $2 \pi / \Delta$. However, in this more general case and for arbitrarily long distances, the transfer of OAM from the plasma rotating vortex to the neutrino beam will eventually be blocked by phase mixing.

\section{Conclusions}

In this work we have considered the orbital angular momentum of a neutrino beam in  a plasma, to our knowledge for the first time. We have studied the interaction of the neutrino beam with  a plasma vortex, and explored the similarities between neutrino and photon dispersion in a plasma.
We have shown that a neutrino beam, propagating in a dense plasma background, can be generally described by a superposition of OAM states, similar to the photon OAM states recently explored in the literature. The existence of plasma vortices introduces coupling between different neutrino OAM states. We have shown that plasma vorticity can be transfered to the neutrino beam. This process can be of particular relevance to the understanding of the astrophysical neutrino sources. Study of total angular momentum conservation, using the existing neutrino detection systems, could eventually lead to the first observation of neutrino OAM states.

In this work, a simplified neutrino description was used. In particular, we have neglected the neutrino mass oscillation process. It is known that in  a dense plasma, these oscillations are resonantly enhanced by the so called MSW effect. This means that the existence of plasma vorticity in the resonant plasma region where the MSW effect takes place will be able to excite OAM states of the different neutrino species. Generalization of the MSW effect, taking the OAM states into account, is in progress and will be the subject of a future work.

\end{document}